\documentclass[twocolumn,
letter
]{jpsj2}
\usepackage[usenames]{color}

\title{
Effects of Electron-Lattice Coupling on Charge Order in $\theta$-(ET)$_2$X
}

\author{
Yasuhiro \textsc{Tanaka$^1$}\thanks{E-mail address:yasuhiro@ims.ac.jp}
and Kenji \textsc{Yonemitsu$^{1,2}$}
}

\inst{
$^1$Institute for Molecular Science, Okazaki, Aichi 444-8585\\
$^2$Graduate University for Advanced Studies, Okazaki, Aichi\ 444-8585
}

\date{\today}

\abst{
Charge ordering accompanied by lattice distortion in quasi-two
dimensional organic conductors $\theta$-(ET)$_2$X (ET=BEDT-TTF) is
studied by using an extended Hubbard model with Peierls-type
electron-lattice couplings within the Hartree-Fock approximation. It is
found that the horizontal-stripe charge-ordered state, which is
experimentally observed in $\theta$-(ET)$_2$RbZn(SCN)$_4$, is stabilized
by the self-consistently determined lattice distortion. Furthermore, in
the presence of the anisotropy in nearest-neighbor Coulomb interactions
$V_{ij}$, the horizontal charge order becomes more stable than any other
charge patterns such as diagonal, vertical and 3-fold-type states. At
finite temperatures, we compare the free energies of various
charge-ordered states and find a first-order transition from
a metallic state with 3-fold charge 
order to the insulating state with the horizontal charge order. The role 
of lattice degrees of freedom in the realization of the horizontal
charge order and the relevance to experiments on $\theta$-(ET)$_2$X are
discussed.
}

\kword{charge order, organic conductor, extended Hubbard model,
electron-lattice coupling, Hartree-Fock approximation}

\begin{document}
\sloppy
\maketitle

Quasi-two-dimensional molecular conductors (ET)$_2$X (ET=BEDT-TTF) 
show a variety of electronic phases at low
temperatures\cite{Ishiguro,Seo_Rev}. Among them,
charge order (CO) phenomena are one of the main subjects and have been 
intensively studied recently. (ET)$_2$X is a member of the
so-called 2:1 salts, which consists of alternating layers of anionic
X$^{-}$ and cationic ET$^{+1/2}$ whose $\pi$-band is $3/4$-filled. 
The variety of their physical properties originates from the spatial
arrangements of ET molecules and strong Coulomb interaction among
$\pi$ electrons.

The experimental observations of CO are made in compounds
such as $\theta$-(ET)$_2$RbZn(SCN)$_4$\cite{Miyagawa,Chiba} and
$\alpha$-(ET)$_2$I$_3$\cite{Takano1,Takano2}.
$\theta$-(ET)$_2$RbZn(SCN)$_4$ shows a
metal-insulator transition at $T=200$ K and a spin gap behavior at low
temperatures\cite{Mori1}. The transition is of first order
accompanied by lattice distortion. CO formation below $T_c$ has been
directly observed in NMR experiments\cite{Miyagawa,Chiba}. Several
experiments\cite{Tajima,Yamamoto,Watanabe1,Watanabe2} such as Raman
scattering\cite{Yamamoto} and X-ray scattering\cite{Watanabe1,Watanabe2} 
measurements indicate that the horizontal-type CO is formed in this
compound.

CO phenomena are considered to be a consequence of strong correlation 
effects among electrons, especially due to the long-range component of
the Coulomb interaction. So far, many theoretical investigations on CO
have been carried out from this point of
view\cite{Seo,Mori2,Kaneko,Clay,Merino,Watanabe3,Kuroki,Hotta}.
For example, Seo studied the extended Hubbard model, that includes both
on-site $(U)$ and intersite $(V)$ Coulomb interactions using the Hartree
approximation and discussed the stability of various stripe CO patterns
in (ET)$_2$X\cite{Seo}. For $\theta$-type salts, the possibility of
CO with long periodicity has been considered within the Hartree
approximation\cite{Kaneko}. Actually, X-ray experiments on
$\theta$-(ET)$_2$RbZn(SCN)$_4$ indicate a short-range CO
with long periodicity in the metallic phase which is different from the
horizontal stripe state at low temperatures\cite{Watanabe1,Watanabe2}. 
A similar charge fluctuation is observed in
$\theta$-(ET)$_2$CsZn(SCN)$_4$, which shows coexisting charge modulations
with different wave vectors without long-range
order\cite{Watanabe4,Nogami}. 

On the other hand, a coupling between electron and lattice degrees of 
freedom also seems to have an important role. In fact, the CO transition
is accompanied by a structural distortion in many $\theta$-type materials
including $\theta$-(ET)$_2$RbZn(SCN)$_4$. Moreover, a particular role of
structural modification at the transition is suggested by a recent 
observation of photoinduced melting of CO in
$\theta$-(ET)$_2$RbZn(SCN)$_4$ and $\alpha$-(ET)$_2$I$_3$\cite{Iwai}.
Several theoretical studies\cite{Seo,Kaneko,Clay} indicate that lattice
effects indeed stabilize the horizontal CO in $\theta$-(ET)$_2$X,
although any electron-lattice coupling which causes structural change is
not explicitly included in the calculations. Thus, it is important to
investigate not only the role of electron-electron interactions but also
lattice effects on CO.

In this paper, we study the CO transition and lattice distortion in 
$\theta$-(ET)$_2$RbZn(SCN)$_4$ by using the extended Hubbard model with 
Peierls-type electron-lattice couplings within the Hartree-Fock
approximation. 
Figures 1(a) and 1(b) show the structures of $\theta$-(ET)$_2$RbZn(SCN)$_4$
in the metallic and insulating phases, which is called $\theta$-type and
$\theta_d$-type, respectively.
At $T>T_c$, the unit cell contains two
molecules and two kinds of transfer integrals, $t_c$ and $t_p$.
On the other hand, six transfer integrals exist in the unit cell with a
doubled $c$-axis at $T<T_c$. Since the displacements of ET
molecules 
and the resulting change in transfer
integrals are rather complicated\cite{Watanabe1}, 
here we study the effects of electron-lattice couplings
which cause the modulations of transfer integrals that are
experimentally observed [Fig. 1(b)], and do not consider any other
electron-lattice couplings. This leads to
three kinds of interactions between electrons and the lattice degrees of
freedom: transfer integrals modulated by $c$- and $a$-axis molecular
translations
and rotation, as deduced from the results of the X-ray
experiment\cite{Watanabe1}. 
For simplicity, these electron-lattice couplings are assumed to be
independent of each other. First, the $c$-axis translation alternates
$t_c$ and gives $t_{c1}$ and
$t_{c2}$ in Fig. 1(b). This is indeed expected since the length of 
the $t_{c1}$ bond increases while that of the $t_{c2}$ bond decreases
through the CO transition\cite{Watanabe1,Def_BL,Watanabe5}. On the other hand,
we observe that the length of the $t_{p1}$ bond decreases while that of
the $t_{p3}$ bond increases\cite{Watanabe1,Watanabe5}, from which the
modulations of $t_{p1}$ and $t_{p3}$ can be regarded as due to the
$a$-axis translation. However, a similar consideration does not hold for
the changes of $t_{p2}$ and $t_{p4}$. 
In fact, the experimental estimation of transfer integrals indicates
that rotational degrees of freedom are important. It shows that the
dependences of the transfer integrals on relative angles (called
elevation angles\cite{Watanabe1}) of ET molecules are large and
allow the transfer integrals $|t_{p2}|$ ($|t_{p4}|$) on the
horizontally connected bonds to uniformly decrease (increase), as can be
seen from Fig. 1(b). In the actual compound, this type of modulation
seems to be important since the horizontal CO is formed by the $t_{p4}$
chains with hole-rich molecules and the $t_{p2}$ chains with hole-poor
molecules. Therefore, in the present
study, we simply introduce such rotational degrees of freedom in order
to take account of the experimentally observed modulations of $t_{p2}$
and $t_{p4}$, which are difficult to understand from a molecular
translation.
\begin{figure}
\begin{center}
\includegraphics[width=8.5cm,clip]{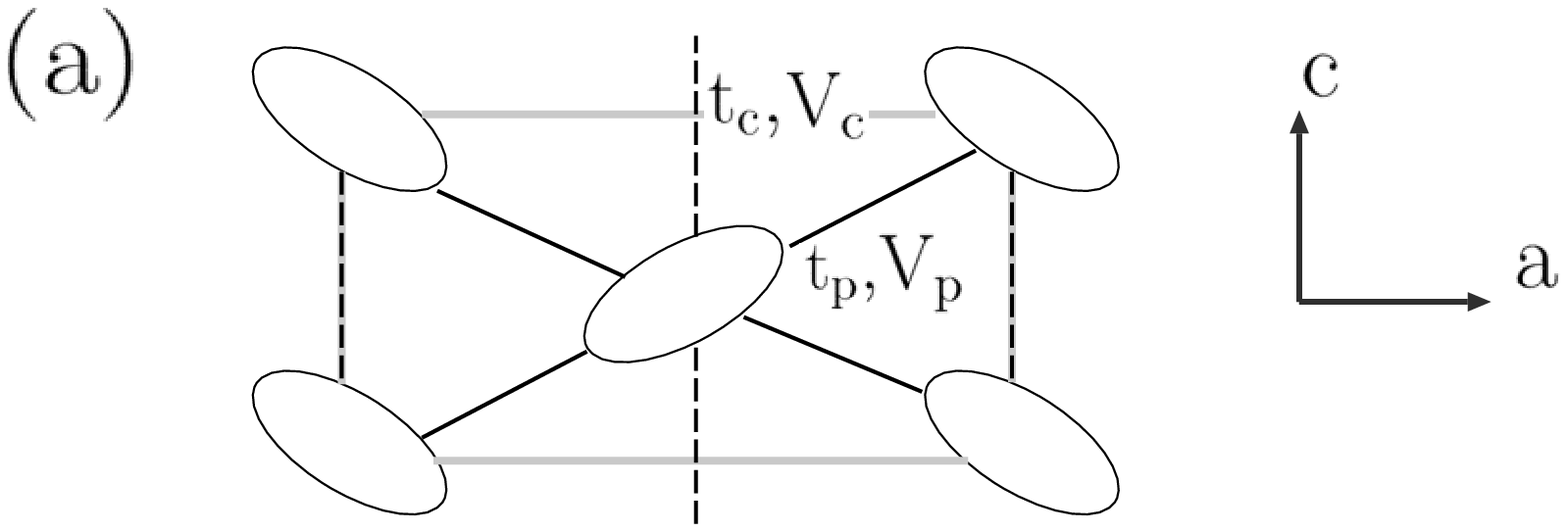}
\end{center}
\vspace{0.5cm}
\begin{center}
\includegraphics[width=8.0cm,clip]{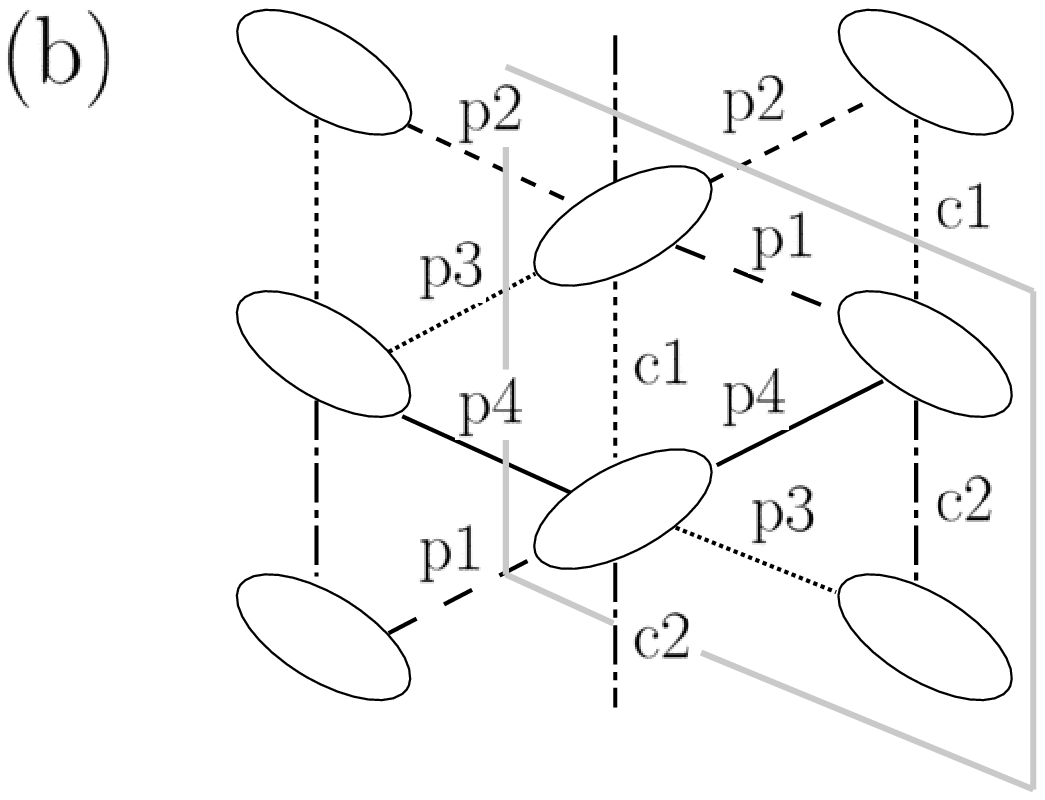}
\end{center}
\caption{Schematic representation of (a) $\theta$-type and (b) $\theta_d$-type
structures. The gray solid lines indicate the unit
 cell. The values of transfer integrals in (b) from the extended
 H$\ddot{\rm{u}}$ckel method are as follows, $t_{c1}=1.5$, $t_{c2}=5.2$,
 $t_{p1}=16.9$,  $t_{p2}=-6.5$, $t_{p3}=2.2$, and
 $t_{p4}=-12.3$($10^{-2}$eV).}
\end{figure}

Our Hamiltonian is then written as 
\begin{equation}
\begin{split}
{\it H}=&\sum_{\langle ij \rangle\sigma}(t_{i,j}+ \alpha_{i,j}u_{i,j})
(c^{\dagger}_{i\sigma}c_{j\sigma}+\rm{h.c})\\
&+U\sum_{i}n_{i\uparrow}n_{i\downarrow}+\sum_{\langle ij \rangle}
 V_{i,j}n_{i}n_{j}+\sum_{\langle ij \rangle}\frac{K_{i,j}}{2}u^{2}_{i,j}\ ,
\end{split}
\end{equation}
where $\langle ij\rangle$ represents the summation over pairs of 
neighboring sites, $c^{\dagger}_{i\sigma}(c_{i\sigma})$ denotes the
creation (annihilation) operator for an electron with spin $\sigma$ at
the $i$th site, $n_{i\sigma}=c^{\dagger}_{i\sigma}c_{i\sigma}$, and
$n_{i}=n_{i\uparrow}+n_{i\downarrow}$. The transfer integral $t_{ij}$
means $t_c$ or $t_p$ in the $\theta$-type structure. The
electron-lattice coupling constant, the lattice translational or
rotational displacement and the elastic constant are denoted by
$\alpha_{i,j}$, $u_{i,j}$, and $K_{i,j}$, respectively.
The electron density is 3/4-filled and we consider nearest-neighbor 
Coulomb interactions $V_c$ for the vertical direction and $V_p$ for
the diagonal direction as shown in Fig. 1(a). For the lattice degrees of
freedom, we further introduce new variables as 
$y_{i,j}=\alpha_{i,j}u_{i,j}$ and $s_{i,j}=\alpha_{i,j}^{2}/K_{i,j}$,
where $s_{i,j}$ is written as $s_{c}$, $s_{a}$ and $s_{\phi}$
for $c$-axis translation, $a$-axis translation and rotation,
respectively, as discussed above. Similarly, we can rewrite $y_{ij}$ by
using the subscripts $c$, $a$ and $\phi$, and as a result
\begin{figure}[h]
\begin{center}
\includegraphics[width=8cm]{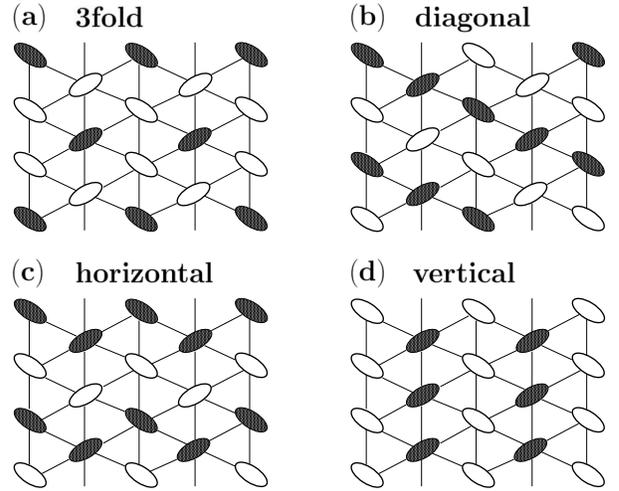}
\end{center}
\caption{Order parameters considered in the Hartree-Fock
 approximation. The hole-rich and -poor sites are represented by the solid
 and open ellipses, respectively.}
\end{figure}
the transfer integrals in the distorted structure are given by
\begin{equation}
\begin{split}
&t_{c1}=t_{c}+y_{c}\ \ , t_{c2}=t_{c}-y_{c}\ ,\\
&t_{p1}=t_{p}+y_{a}\ \ , t_{p2}=t_{p}-y_{\phi}\ ,\\ 
&t_{p3}=t_{p}-y_{a}\ \ , t_{p4}=t_{p}+y_{\phi}\ .
\end{split}
\end{equation}
Note that the signs in eq. (2) are chosen so that the resulting 
transfer integrals correspond with the experimental ones if 
$y_{l}>0$ for $l=c,\ a$ and $\phi$\cite{Note}.

We apply the Hartree-Fock approximation,
\begin{equation}
\begin{split}
n_{i\sigma}n_{j\sigma^{\prime}}\rightarrow
&\langle n_{i\sigma}\rangle n_{j\sigma^{\prime}}+
n_{i\sigma}\langle n_{j\sigma^{\prime}}\rangle -\langle
 n_{i\sigma}\rangle \langle n_{j\sigma^{\prime}}\rangle \\
&-\langle c_{i\sigma}^{\dagger}c_{j\sigma^{\prime}}\rangle
c_{j\sigma^{\prime}}^{\dagger}c_{i\sigma}-c_{i\sigma}^{\dagger}c_{j\sigma^{\prime}}
\langle c_{j\sigma^{\prime}}^{\dagger}c_{i\sigma}\rangle \\
&+\langle c_{i\sigma}^{\dagger}c_{j\sigma^{\prime}}\rangle \langle
 c_{j\sigma^{\prime}}^{\dagger}c_{i\sigma}\rangle\ ,
\end{split}
\end{equation}
to eq. (1) and diagonalize the obtained Hamiltonian in $k$-space by
assuming the unit cell of each mean-field order parameter. We use four
types of CO order parameters with respect to charge degrees of freedom,
which are shown in Fig. 2.
As for the spin degrees of
freedom, we consider three spin configurations in each stripe-type CO
which are identical to those of ref. 12. For the 3-fold CO, spin
alternation between the hole-rich and -poor sites is
considered. The ground-state energy is calculated by solving the 
mean-field equation self-consistently with the lattice
displacements, which are determined by the Hellmann-Feynman theorem
$\left \langle \frac{\partial H}{\partial y_{l}}\right \rangle = 0$,
where $y_l$ means $y_c$, $y_{a}$ or $y_{\phi}$. The energy per site
is given by
\begin{equation} 
\begin{split}
E &=
 \frac{1}{N}\Bigl(\sum_{l\bf{k}\sigma}E_{l\bf{k}\sigma}n_{F}(E_{l\bf{k}\sigma})-
U\sum_{i}\langle n_{i\uparrow}\rangle \langle n_{i\downarrow}\rangle \\
&-\sum_{\langle ij\rangle}V_{ij}\langle n_{i}\rangle \langle 
n_{j}\rangle +\sum_{\langle ij\rangle \sigma}V_{ij}\langle
 c_{i\sigma}^{\dagger}c_{j\sigma}\rangle \langle
 c_{j\sigma}^{\dagger}c_{i\sigma}\rangle \\
&+\sum_{\langle ij\rangle}\frac{y_{ij}^{2}}{2s_{ij}}\Bigr)\ ,
\end{split}
\end{equation}
where {\it l}, $E_{l\bf{k}\sigma}$ and $n_{F}$ are the band index, the
energy eigenvalue of the mean-field Hamiltonian and the Fermi distribution
function, respectively. $N$ is the total number of sites. In the
following, we set $t_p=0.1$ eV, $t_c=-0.04$ eV, and $U=0.7$ eV. The ratio
$V_c/U$ is fixed at 0.35 and the anisotropy in nearest-neighbor Coulomb
interactions $V_p/V_c$ is treated as a parameter.

The ground-state energies of various CO patterns per site as a
function of $V_p/V_c$ are compared in Fig. 3, where the energy of the
3-fold CO is set at zero. 
We have shown only the lowest-energy state of each CO pattern with
different spin configurations. 
In the absence of an electron-lattice coupling, the 3-fold CO with
a ferrimagnetic spin configuration is the most favorable in the nearly isotropic
region, i.e., $V_p/V_c\sim 1$. On the other hand, the diagonal CO whose
spin configuration is antiferromagnetic along the stripe and between
stripes on the $c$-axis is stable when $V_p/V_c$ is small. These
features are consistent with the previous study\cite{Kaneko}. For
the horizontal CO, we plotted the energy of the state which is
antiferromagnetic along the stripe and 
ferromagnetic between stripes on the $c$-axis. Note that the state
which is antiferromagnetic on the $c$-axis has a close energy and is
nearly degenerate with the above state. As can be seen from Fig. 3,
there is no region where the horizontal CO has the lowest energy in the
absence of an electron-lattice coupling.

\begin{figure}
\begin{center}
\includegraphics[width=8.5cm]{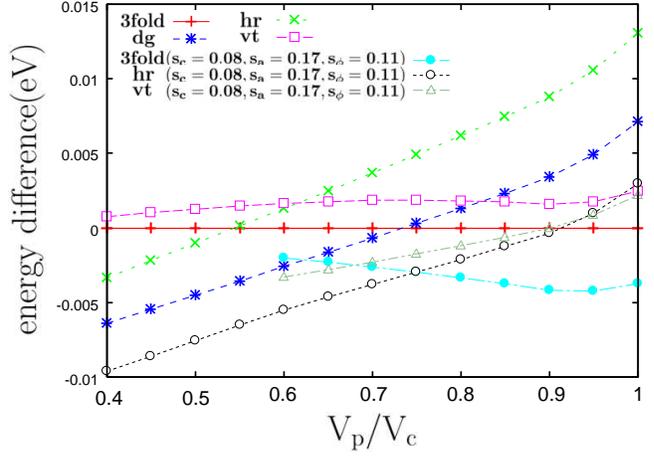}
\end{center}
\caption{\hspace{-0.2cm}(Color online)\ Relative energies as a function of anisotropy
 $V_p/V_c$, where the energy of the 3-fold state is chosen to be zero. dg,
 hr and vt are abbreviations of diagonal, horizontal and vertical COs,
 respectively.}
\vspace*{-0.2cm}
\end{figure}
\begin{figure}
\begin{center}
\includegraphics[width=7.6cm]{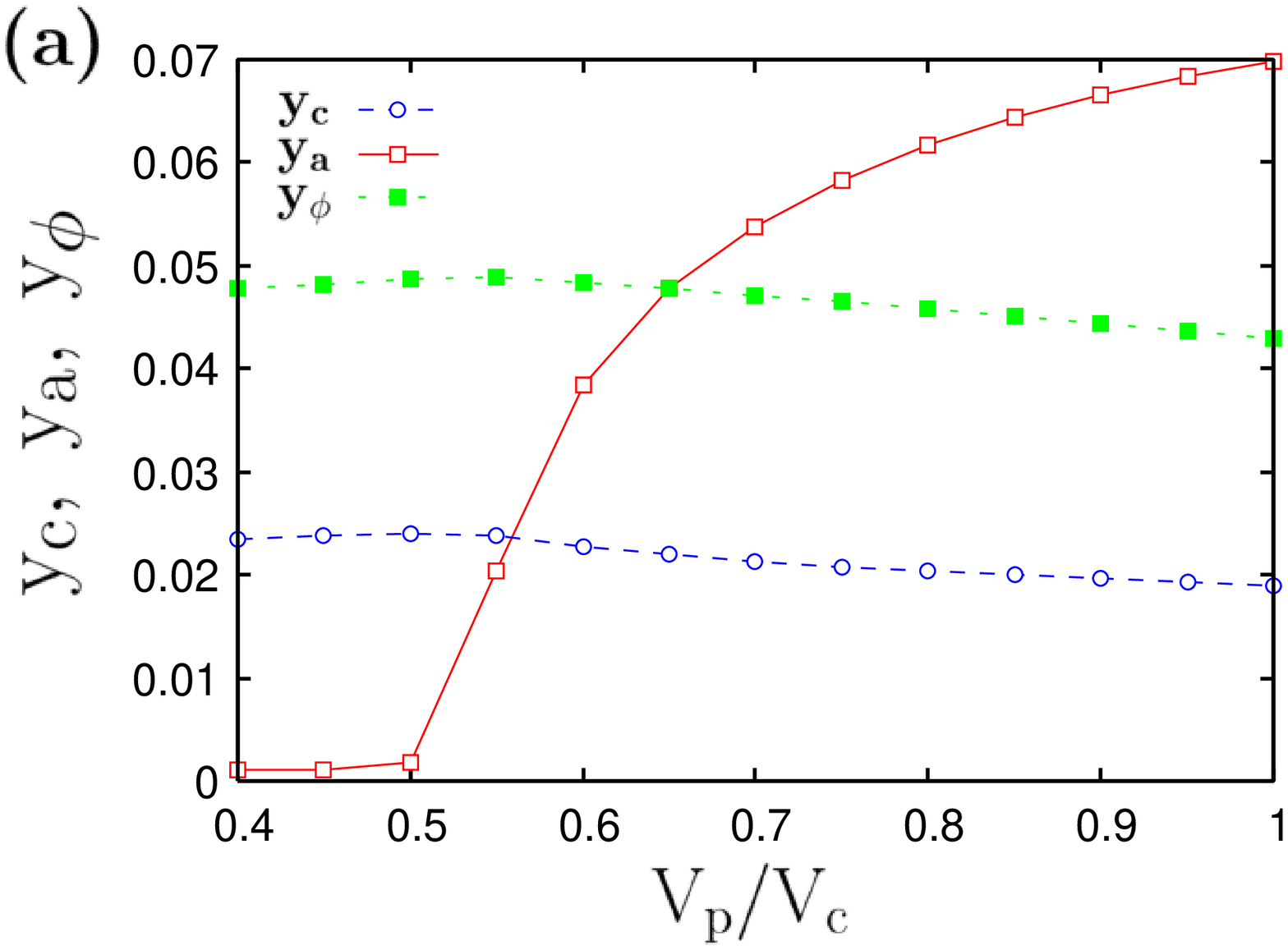}
\end{center}
\begin{center}
\includegraphics[width=7.6cm]{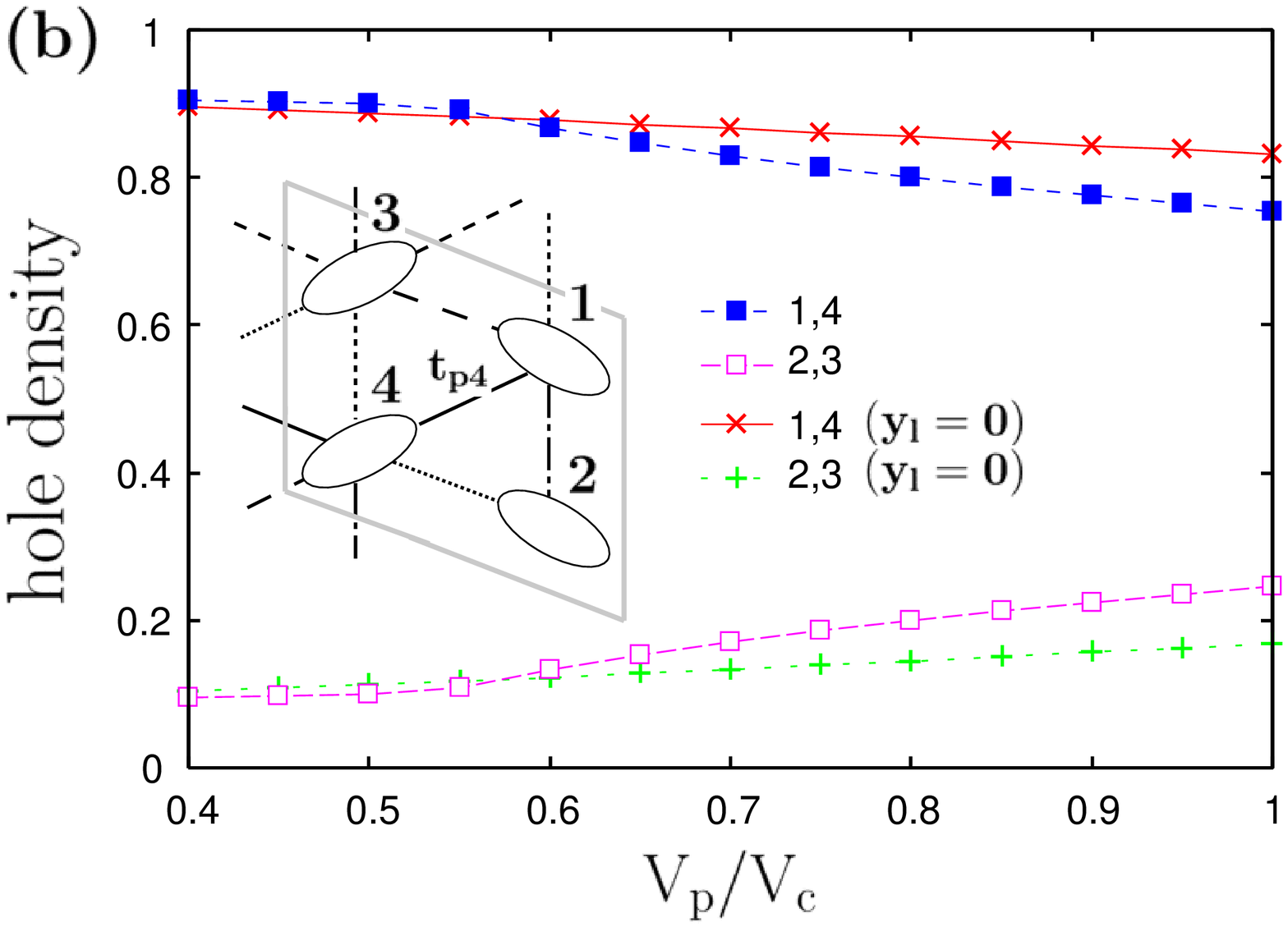}
\end{center}
\caption{\hspace{-0.2cm}(Color online)\ (a) Modulations of transfer
 integrals and (b) hole density at each site in the horizontal CO. In
 (b), the hole densities in the absence of an electron-lattice coupling
 are also shown.}
\end{figure}
However, in the presence of the electron-lattice couplings, the
horizontal CO becomes more stable owing to the lattice distortions. The
values of the
electron-lattice couplings are chosen at $s_c=0.08$, $s_a=0.17$ and 
$s_{\phi}=0.11$ to obtain realistic values of lattice displacements. The
horizontal CO has hole-rich sites on the $t_{p4}$ chains, which is
consistent
with the experiments. The energy gain mainly comes from the difference
between $t_{p2}$ and $t_{p4}$. This is reasonable since the horizontal
CO can be stabilized by the exchange coupling between neighboring spins
on the stripes.
Although the diagonal CO is not affected by any electron-lattice
coupling, the vertical and 3-fold COs also have energy gain from
the lattice modulation. The energy of the vertical CO is lowered by
$s_{a}$. 
On the other hand, that of the 3-fold CO is lowered by $s_{c}$ and
$s_{\phi}$. In this state, there is a weak horizontal charge modulation 
caused by the lattice distortion in the background of the 3-fold CO.
Note that the 3-fold CO is metallic even if the transfer integrals are
modulated, whereas the horizontal CO is insulating. As a result, the
horizontal CO with lattice distortion becomes stable for
$V_p/V_c<0.75$, while the 3-fold CO is favorable for $V_p/V_c>0.75$, as
shown in Fig. 3.

In Fig. 4, we show the modulations of the transfer integrals and the
hole density at each site in the case of the horizontal CO. Although
$y_{c}$ and $y_{\phi}$ distortions do not depend so much on $V_p/V_c$,
the $y_{a}$ distortion increases with $V_p/V_c$. It becomes largest for
$V_p/V_c\ge 0.7$. In fact, the former two electron-lattice couplings
favor the horizontal CO while the latter tends to decrease the order
parameter as seen from Fig. 4(b), although the energy is
lowered. Experimentally, the difference between
$t_{p1}$ and $t_{p3}$ is the largest while that between $t_{c1}$ and $t_{c2}$
is the smallest. Our result seems to be consistent with the experimental
one for $V_p/V_c\sim 0.7$.
The detailed role of each electron-lattice coupling on the horizontal CO
is discussed elsewhere\cite{Tanaka}.

Next, we consider the stability of these COs at finite temperatures by
calculating the free energy within the Hartree-Fock approximation. The
free energy per site is written as
\begin{equation} 
\begin{split}
F &=
 \frac{1}{N}\Bigl(\mu N_{tot}-\frac{1}{\beta}\sum_{l\bf{k}\sigma}\ln
 (1+\exp \{-\beta (E_{l\bf{k}\sigma}-\mu)\}) \\
&-U\sum_{i}\langle n_{i\uparrow}\rangle \langle n_{i\downarrow}\rangle 
-\sum_{\langle ij\rangle}V_{ij}\langle n_{i}\rangle \langle n_{j}\rangle
 \\
&+\sum_{\langle ij\rangle \sigma}V_{ij}\langle
 c_{i\sigma}^{\dagger}c_{j\sigma}\rangle \langle
 c_{j\sigma}^{\dagger}c_{i\sigma}\rangle
 +\sum_{\langle ij\rangle}\frac{y_{ij}^{2}}{2s_{ij}}\Bigr)\ ,
\end{split}
\end{equation}
where $\mu$, $N_{tot}$ and $\beta$ are the chemical potential, the total
number of electrons and the inverse temperature, respectively. The 
phase diagram on the $(T,V_p/V_c)$ plane, which is obtained by comparing
the free energies of different CO patterns, is shown in Fig. 5. The
values of the electron-lattice couplings are the same as those used for
$T=0$.
\begin{figure}[h]
\begin{center}
\includegraphics[width=7.5cm]{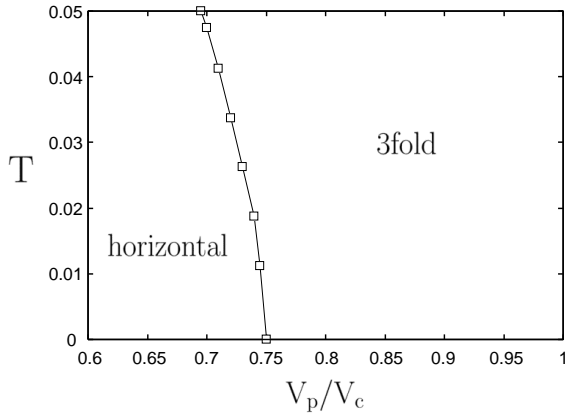}
\end{center}
\caption{Phase diagram on $(T,V_p/V_c)$ plane in the presence of
 electron-lattice coupling.}
\end{figure}
For $V_p/V_c\sim 1$, the 3-fold state with lattice distortion has the 
lowest free energy for a wide temperature range. On the other hand,
the horizontal CO is the most stable for $V_p/V_c<0.7$. There 
is a first-order metal-insulator transition from the 3-fold CO
to the horizontal CO near $V_p/V_c=0.7$. For the parameters we
used, the normal metallic state with a uniform charge density always has
a higher free energy than the 3-fold CO because of the large Coulomb
interactions. We note that if we choose smaller values of $U$ and
$V_{ij}$, the diagonal CO, which is undistorted even
with electron-lattice couplings, is more stable than the horizontal
CO. Therefore, large Coulomb interactions
seem to be important to stabilize the horizontal CO with realistic
values of the lattice distortions. In fact, the energy gain of the
horizontal CO due to the lattice distortions can be understood by the
perturbational calculations from the strong couping limit, i.e.,
$t_{ij}=0$\cite{Miyashita}.

Finally, we discuss the relevance of the results to the experiments
and relations to other theoretical studies. The stabilization of
the horizontal CO due to the lattice distortion is consistent with the 
experiments on $\theta$-(ET)$_2$RbZn(SCN)$_4$. Since the horizontal
CO does not become the ground state without electron-lattice
coupling, the effects of the lattice distortion are considered to be
crucial in realizing the horizontal CO in the present model.
This result is also qualitatively consistent with the recent
exact-diagonalization study\cite{Miyashita} for eq. (1) on small
clusters.
Moreover, the first-order metal-insulator transition at a finite
temperature can be related to the experimental results of this compound,
although the obtained wave vector of the charge modulation at high
temperatures is different from that of the experiments in the
metallic phase. It has recently been pointed out that longer range
than the nearest-neighbor Coulomb interactions can reproduce the
experimental observation\cite{Kuroki}. As for the spin degrees of
freedom, both the 3-fold and horizontal COs in our Hartree-Fock
calculation have spin orders which have not been observed in the
experiments. It is considered that the effect of quantum fluctuation
is necessary in discussing the behavior of the spin degrees of
freedom\cite{Kaneko}.

The previous estimations of the intersite Coulomb interactions $V_p$
and $V_c$ show that these values are comparable, $V_p/V_c\sim
1$\cite{Mori2}, where the 3-fold CO is the most stable in our calculation.
A variational Monte Carlo study\cite{Watanabe3} in the absence of 
an electron-lattice coupling also shows that the 3-fold CO is stable for 
$V_p/V_c\sim 1$.
According to the recent exact-diagonalization study\cite{Miyashita}, the
horizontal CO with lattice distortion becomes more stable even at
$V_p/V_c\sim 1$ if we take account of quantum fluctuations that are
neglected in the Hartree-Fock approximation. 
An exact-diagonalization study\cite{Clay} also indicates that the
Holstein-type electron-lattice coupling stabilizes the horizontal CO.

At the nearly isotropic region $V_p/V_c\sim 1$, we find that the 3-fold
state with a coexisting weak horizontal charge modulation is stable. This
result can be related to the X-ray experiments on
$\theta$-(ET)$_2$CsZn(SCN)$_4$\cite{Watanabe4,Nogami}, which shows two
types of COs coexisting as short-range fluctuations. 
Although the present Hartree-Fock calculation gives a long-range CO, it
is natural to expect that the effect of fluctuations can destroy the
long-range order and results in a state such that two types of COs
coexist as short-range fluctuations.

In summary, we investigated the effects of Peierls-type electron-lattice
couplings on the CO in $\theta$-(ET)$_2$X by using the extended Hubbard
model through the Hartree-Fock approximation. We found that the
horizontal stripe CO which is observed in the experiments is stabilized
by the lattice distortion. Moreover, at finite temperatures, there is a
first-order metal-insulator transition in the presence of the anisotropy
in $V_{ij}$, which can be related to the phase transition in
$\theta$-(ET)$_2$RbZn(SCN)$_4$. These results show that the lattice
effect plays an important role on the CO phenomena in $\theta$-(ET)$_2$X. 

\section*{Acknowledgment}
The authors would like to thank H. Seo and S. Miyashita for helpful
discussions. This work was supported by Grants-in-Aid and the Next
Generation Super Computing Project, Nanoscience Program, from the
Ministry of Education, Culture, Sports, Science and Technology, Japan.


\begin{thebibliography}{99}
\bibitem{Note}
In the present study, we choose $t_p>0$ and $t_c<0$. In this choice, a
	correspondence with the experimental values of transfer
	integrals in the low-temperature phase is obtained since we can
	change the signs of $t_{c1}$, $t_{c2}$, $t_{p2}$ and $t_{p4}$
	simultaneously by the redefinition of electron operators.
\end{thebibliography}
\end{document}